\documentclass[acmlarge]{acmart}

\AtBeginDocument{%
  \providecommand\BibTeX{{%
    \normalfont B\kern-0.5em{\scshape i\kern-0.25em b}\kern-0.8em\TeX}}}

\setcopyright{acmcopyright}
\copyrightyear{}
\acmYear{}
\acmDOI{}

\usepackage{amsmath}
\usepackage{url}
\usepackage{longtable}
\usepackage{booktabs}
\usepackage{multicol}
\usepackage{multirow}
\usepackage{lineno}

\usepackage{tikz}
\usepackage{tikzscale}
\usepackage{xspace}
\usepackage{pgfplots}
\usepackage{multirow}
\usepackage{textcomp}
\usepackage{graphicx}
\usepackage{caption}
\usepackage{subcaption}
\usepackage{float}
\usepackage{algorithm}
\usepackage{algpseudocode}
\usepackage[colorinlistoftodos]{todonotes}

\usepackage{hyperref}

\theoremstyle{definition}

\acmConference{International Conference on Big Data Research}
\acmYear{2021}

\begin{document}

\title{Effective and scalable clustering of SARS-CoV-2 sequences}

\author{Sarwan Ali}
\email{sali85@student.gsu.edu}
\affiliation{%
  \institution{Georgia State University, Atlanta GA}
  \country{USA}
}

\author{Tamkanat-E-Ali}
\email{20100159@lums.edu.pk}
\affiliation{%
  \institution{Lahore University of Management Sciences, Lahore}
  \country{Pakistan}
}

\author{Muhammad Asad Khan}
\email{asadkhan@hu.edu.pk}
\affiliation{%
  \institution{Hazara University, Mansehra}
  \country{Pakistan}
}

\author{Imdadullah Khan}
\email{imdad.khan@lums.edu.pk}
\affiliation{%
  \institution{Lahore University of Management Sciences, Lahore}
  \country{Pakistan}
}

\author{Murray Patterson}
\email{mpatterson30@gsu.edu}
\affiliation{%
  \institution{Georgia State University, Atlanta GA}
  \country{USA}
}

\renewcommand{\shortauthors}{S. Ali et al.}


\begin{abstract}

SARS-CoV-2, like any other virus, continues to mutate as it spreads,
according to an evolutionary process.  Unlike any other virus, the
number of currently available sequences of SARS-CoV-2 in public
databases such as GISAID is already several million.  This amount of
data has the potential to uncover the evolutionary dynamics of a virus
like never before.  However, a million is already several orders of
magnitude beyond what can be processed by the traditional methods
designed to reconstruct a virus's evolutionary history, such as those
that build a phylogenetic tree.  Hence, new and scalable methods will
need to be devised in order to make use of the ever increasing number
of viral sequences being collected.

Since identifying variants is an important part of understanding the
evolution of a virus, in this paper, we propose an approach based on
clustering sequences to identify the current major SARS-CoV-2
variants.  Using a $k$-mer based feature vector generation and
efficient feature selection methods, our approach is effective in
identifying variants, as well as being efficient and scalable to
millions of sequences.  Such a clustering method allows us to show the
relative proportion of each variant over time, giving the rate of
spread of each variant in different locations --- something which is
important for vaccine development and distribution.  We also compute
the importance of each amino acid position of the spike protein in
identifying a given variant in terms of information gain.  Positions
of high variant-specific importance tend to agree with those reported
by the USA's Centers for Disease Control and Prevention (CDC), further
demonstrating our approach.

\end{abstract}



\maketitle

\section{Introduction}

The current global COVID-19 pandemic is caused by SARS-CoV-2 (severe
acute respiratory syndrome coronavirus 2) and its variants.
SARS-CoV-2 was first detected in Wuhan, China by the end of 2019 (12
December 2019)~\cite{wu2020new}.  Because SARS-CoV-2 is easily
transmitted from one person to another, it spread (exponentially)
quickly throughout the globe, developing variants such as Alpha and
now Delta, which have greatly accelerated this spread.  As of July
2021, more than 200 million COVID-19 cases have been reported in 221
countries, with over 4 million deaths~\cite{covid_statistics_cnn}.

From the perspective of molecular evolution, SARS-CoV-2 continues to
diverge as it spreads, developing new variants, like any other virus.
Hence, methods for building phylogentic trees~\cite{hadfield2018a} ---
which have been successful for Ebola, Zika and the Avian Flu --- seem
apt, because they provide a very complete and detailed evolutionary
history of a virus based on molecular sequencing data.  The only
problem is that such tree building methods can typically process
thousands of sequences, while the number of sequences of SARS-CoV-2,
available in public databases such as
GISAID~\cite{gisaid_website_url}, is already several million.
Approaches which first divide the dataset into smaller
subsets~\cite{du2021establishment}, or use
parallelization~\cite{minh_2020_iqtree21}, have allowed the building of
trees to scale to tens of thousands of SARS-CoV-2 sequences, however
this is still far from closing the gap.

The number of available SARS-CoV-2 sequences forces us to rethink the
current methodology used to study the evolutionary dynamics of
viruses, and this number will only increase in the coming
decades~\cite{stephens-2015-genomical}.  While building a phylogenetic
tree directly on the millions of SARS-CoV-2 sequences seems out of
reach, variants of the virus would appear as major clades of the
hypothetical SARS-CoV-2 phylogenetic tree.  Since the major clades
form natural \emph{clusters} of the sequences, a completely orthogonal
approach based on clustering the sequences directly could shed light
on this important part of the evolutionary dynamics of SARS-CoV-2 (or
viruses in general).  Identifying variants has many practical
application as well, guiding vaccine design and distribution
decisions, as well as informing strategies on how to monitor and
prevent future outbreaks.  A major advantage of clustering is that it
is a widely studied field for which there are techniques which can
scale to millions and billions of objects.  It remains to decide which
(parts of) SARS-CoV-2 sequences to choose, and the best feature vector
representation to use for the downstream clustering, which we describe
below.



\begin{figure}[!ht]
  \centering
  \includegraphics[scale=0.4,page=1]{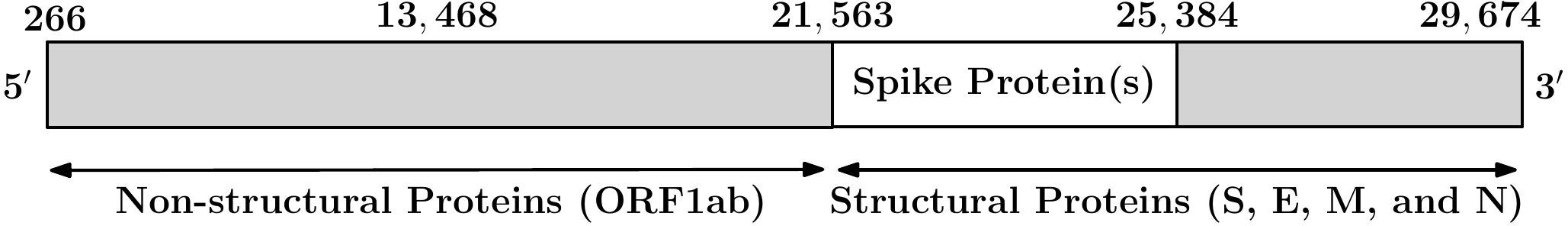}
  \caption{The SARS-CoV-2 genome is around 29--30kb encoding
    structural and non-structural proteins.  The four structural
    proteins include spike, envelope, membrane, and nucleocapsid.  The
    spike region comprises $25,384 - 21,563 = 3821$ nucleotides, which
    code for a sequence of 3821 (+ 1 stop codon *) $/ 3 = 1274$ amino
    acids.}
  \label{fig_spike_seq_example}
\end{figure}

It is common knowledge that the majority of the variation in the
SARS-CoV-2 genome takes place in the spike
region~\cite{kuzmin2020machine}.  Part of this is because the spike
region encodes an important function of the virus, namely the spike
protein.  For this reason, we focus not only on the spike region of
the genome, but on the amino acid (protein) sequence encoded by this
region (see Figure~\ref{fig_spike_seq_example}).  This allows us to
work with a much more compact representation (avoiding the curse of
dimensionality).  Most of the machine learning (ML) based methods
(classification, clustering, etc.) take numerical (real-valued)
vectors as an input.  Therefore, we cannot work directly with these
amino acid sequences if we want to make use of ML based algorithms.
Since the order of the amino acids within each sequence also matters,
we cannot simply convert the alphabets (amino acids) into numeric
characters and apply ML algorithms.  A popular approach to preserve
the order of the sequences in converting the alphabets vector into
numeric vector is to use the one-hot encoding
approach~\cite{kuzmin2020machine}.  In theory, while the one-hot
encoding preserves the order of amino acid sequences, the preserved
order is of no use while computing pairwise distance (e.g., euclidean
distance)~\cite{ali2021k}.  To deal with this problem, each sequence
is first converted into length $k$ substrings (called $k$-mers).
These $k$-mers successfully preserve the order of each sequence, that
can be crucial to the performance of the classification/clustering
tasks~\cite{farhan2017efficient,ali2021k}.

In this paper, we propose a method to quickly and effectively cluster
the spike amino acid sequences of SARS-CoV-2.  We show that our method
performs substantially better than the baseline approach, and
successfully clusters the variants into unique clusters with high F1
score.  Our contributions in this paper are the following:
\begin{enumerate}
\item We propose a method based on $k$-mers for efficient sequence
  clustering, and show that the quality of clusters as a result of our
  methods is very high.
\item We use this to analyze the spread pattern of different variants
  over time, which can guide vaccination design and distribution
  strategies.
\item We compute the importance of each amino acid position in
  distinguishing a variant, and show that this corresponds to previous
  studies.
\end{enumerate}

The rest of the paper is organized as follows:
Section~\ref{sec_related_work} contains related work.  Our proposed
approach is detailed in Section~\ref{sec_proposed_approach}.  A
description of the datasets used are given in
Section~\ref{sec_experimental_setup}.  We provide a detailed
discussion about the results in
Section~\ref{sec_results_and_discussion} and then we finally conclude
our paper in Section~\ref{sec_conclusion}.

\section{Related Work}
\label{sec_related_work}

Sequence classification and clustering are popular
approaches~\cite{Krishnan2021PredictingVaccineHesitancy,ali2021k}, most methods
requiring the sequences to be
aligned~\cite{Dwivedi2012ClassificationOH}.  Such aligned sequences
are used to design fixed-length feature vector representations that
can be input to classical machine learning (ML) based algorithms.
Authors in~\cite{Chowdhury2017MultipleSequences} compute pairwise
local and global alignment similarity scores between sequences so that different types of analysis can be performed.  Another approach to analyzing the sequences is using heuristic methods.  However, these methods require a number of ad-hoc settings (e.g., a penalty for gaps and substitutions).  Moreover, these methods may not perform well on specific regions of the genome with high variance. Feature vector based representation has been successfully used in literature for other applications such as missing values prediction in graphs~\cite{ali2021predicting}, sequence analysis in  texts~\cite{Shakeel2020LanguageIndependent,Shakeel2020Multi,Shakeel2019MultiBilingual}, biology~\cite{leslie2002mismatch,farhan2017efficient,Kuksa_SequenceKernel}, graph analytics~\cite{hassan2020estimating,Hassan2021Computing}, classification of electroencephalography and
electromyography sequences~\cite{atzori2014electromyography,ullah2020effect}, detecting security attacks in networks~\cite{Ali2019Detecting1}, and electricity consumption in smart grids~\cite{ali2019short,Ali2020ShortTerm}. It is also important to study the conditional dependencies between variables so that their importance can be analysed~\cite{ali2021cache}. Studying the relationships between amino acid positions (attributes) and the class labels (variants) can help us understand which amino acids play a key role in developing specific COVID-19 variants. Information related to mutations and variants could also help the relevant public health departments to study the transmission patterns of different variants, which could help to prevent rapid
spread of the underlying virus~\cite{Ahmad2016AusDM,ahmad2017spectral,Tariq2017Scalable,AHMAD2020Combinatorial}.


To address problems that arise in the alignment-based approaches,
various alignment-free methods have been
proposed~\cite{farhan2017efficient,Chang2014ANA}.  One such method
uses $k$-mers, which preserve the order of the sequences without
relying on alignment.  Authors
in~\cite{Blaisdell1986AMeasureOfSimilarity} use frequency vectors
generated from the $k$-mers for phylogenetic applications.  These
methods are successful in constructing accurate phylogenetic trees
from several (coding and non-coding) DNA sequences.  To measure the
(pairwise) similarity between these frequency vectors, different
distance measures are used~\cite{Zielezinski2017AlignmentfreeSC}.
A recent article~\cite{melnyk2020clustering} proposes a clustering
approach for SARS-CoV-2 subtypes identification, however they cluster
full-length nucleotide sequences, and use a very different
representation and clustering technique.

\section{Proposed Approach}
\label{sec_proposed_approach}

In this section, we first detail the computation of $k$-mers and the
generation of their frequency vectors.  We then show how we apply
feature selection methods on these frequency vectors, followed by
clustering.

\subsection{k-mers Computation}

\begin{figure}[!ht]
  \centering \includegraphics[scale = 0.35] {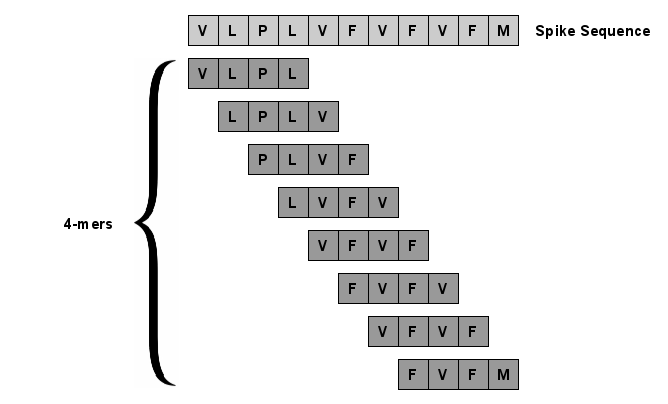}
  \caption{Example of 4-mers of the amino acid sequence
    ``VLPLVFVFVFM''.}
  \label{fig_k_mer_demo}
\end{figure}

The first step of our pipeline is to compute all $k$-mers of each
sequence to map it to a fixed length vector, while allowing its order
to be preserved.  Given a sequence, the total number of $k$-mers that
can be generated is $N - k + 1$, where $N$ is the total number of
elements in the sequence ($1274$ amino acids in our case), and $k$ is
a user-defined parameter for the size of each mer.  See
Figure~\ref{fig_k_mer_demo} for an example.  In our experiments we use
$k = 3$ --- decided using a standard validation set
approach~\cite{validationSetApproach}.

After computing the $k$-mers for each sequence, the next step is to
generate the numerical representation of these vectors.  For this
purpose, we design frequency vectors that contain the counts of each
$k$-mer in the corresponding sequence.

\subsection{Frequency Vectors Generation}

Each sequence $X$ is over an alphabet $\Sigma$ (amino acids in our
case).  Since each frequency vector $\Phi_k(X)$ must have the same
fixed length, this length is $|\Sigma|^k$, the number of possible
$k$-mers of $X$.  In our experiments, the length of each frequency
vector is $4977$.

\subsection{Feature Selection}

After generating the $k$-mers based frequency vector $\Phi_k(X)$ for
each sequence $X$, the next step is to select the most important
features in the feature vectors.  For this purpose, we use the
following two supervised feature selection methods.

\subsubsection{Ridge Regression~\cite{hoerl1975ridge}}

Ridge Regression (RR) is a popular method for feature selection, which is used to address the colinearity problem that often arises in multiple linear regression~\cite{ali2021simpler}.  RR works by
introducing a bias term to increase the bias and hence improve the
variance (i.e., generalization capability of the model).  RR changes
the slope of the (regression) line and tries to make it more
horizontal.  RR is beneficial for feature selection because it gives
insights on which independent variables are not informative (can
reduce the slope close to zero).  We can eliminate some of these
useless independent variables (hence reduce the dimensions of the
data).  The objective function of RR is $min(\text{Sum of square
  residuals} + \alpha \times \text{slope}^2)$, where $\alpha \times
{slope}^2$ is a penalty term.  After applying RR, we selected $1242$
features out of $4977$.

\subsubsection{Lasso Regression~\cite{tibshirani1996regression}}

The Lasso Regression (LR) is another supervised feature selection
algorithm similar to RR~\cite{ranstam2018lasso}.  The only difference
is that it can reduce the slope exactly (i.e., $slope = 0$) rather
than close to zero (as is done by RR).  The objective function of LR
is $min(\text{Sum of square residuals} + \alpha \times \vert
\text{slope}\vert)$, where the penalty term $\alpha \times \vert
\text{slope}\vert$ is now in terms of the cardinality of the slope.
This helps to reduce the slope of useless variables exactly to zero.
After applying lasso regression, we selected $964$ features of $4977$.


\subsection{Clustering using K-means}

After generating $k$-mers, frequency vectors, and feature selection,
we use the $K$-means clustering algorithm (with default parameters) to
cluster the data.  Since we have $5$ variants in our data, we used
$K=5$ for in $K$-means (see Section~\ref{sec_optimal_num_of_clust} for
more details regarding the optimal number of clusters).


\section{Experimental Setup}
\label{sec_experimental_setup}

In this section, we first report information related to the dataset.
Then we try to find any natural (hidden) clustering in the data by
using the t-distributed stochastic neighbor embedding (t-SNE)
approach~\cite{van2008visualizing}.  For the baseline, we use simple
$K$-means clustering algorithm without first applying any feature
selection on the frequency vectors.  To measure the quality of
clustering (without and with feature selection methods), we use the
weighted F1 score.  All experiments are performed on a Core i5 system
running the Windows operating system, 32GB memory, and a 2.4 GHz
processor.  Implementation of the algorithms is done in Python, and
the code is available
online\footnote{\url{https://github.com/sarwanpasha/Visualization_covid_data}}.

\subsection{Dataset Statistics}

We used the (aligned) amino acid sequences corresponding to the spike
protein from the largest known database of SARS-CoV-2 sequences,
GISAID\footnote{\url{https://www.gisaid.org/}}.  In our dataset, we
have $5$ most common variants known to date.  More information related
to the dataset is given in Table~\ref{tbl_variant_information}.  After
preprocessing (removing missing values and truncated sequences), we
end up with $62,657$ amino acid sequences.

\begin{table}[ht!]
  \centering
  \begin{tabular}{p{1.2cm}llp{2.5cm} | p{1.5cm}}
    \hline
    
      Pango Lineage & Region & Labels &
	Num. Mutations S-gene/Genome &  Num. of sequences\\
      \hline	\hline	
      B.1.1.7 & UK~\cite{galloway2021emergence} &  Alpha & 8/17 & \hskip.1in 13966\\
      B.1.351  & South Africa~\cite{galloway2021emergence}  &  Beta & 9/21& \hskip.1in 1727\\
      B.1.617.2  & India~\cite{yadav2021neutralization}  &  Delta &  8/17  & \hskip.1in 7551\\
      P.1  &  Brazil~\cite{naveca2021phylogenetic} &  Gamma &  10/21 & \hskip.1in 26629\\
      B.1.427   & California~\cite{zhang2021emergence}  & Epsilon  &  3/5 & \hskip.1in 12784\\
      \hline
  \end{tabular}
  \caption{Variants information and distribution in the dataset. The
    S/Gen. column represents number of mutations on the S gene /
    entire genome.  Total number of amino acid sequences in our
    dataset is $62,657$.}
  \label{tbl_variant_information}
\end{table}

\subsection{Data Visualization}

\begin{figure}[ht!]
  \centering
  \begin{subfigure}{.5\textwidth}
    \centering
    \includegraphics[scale = 0.35]{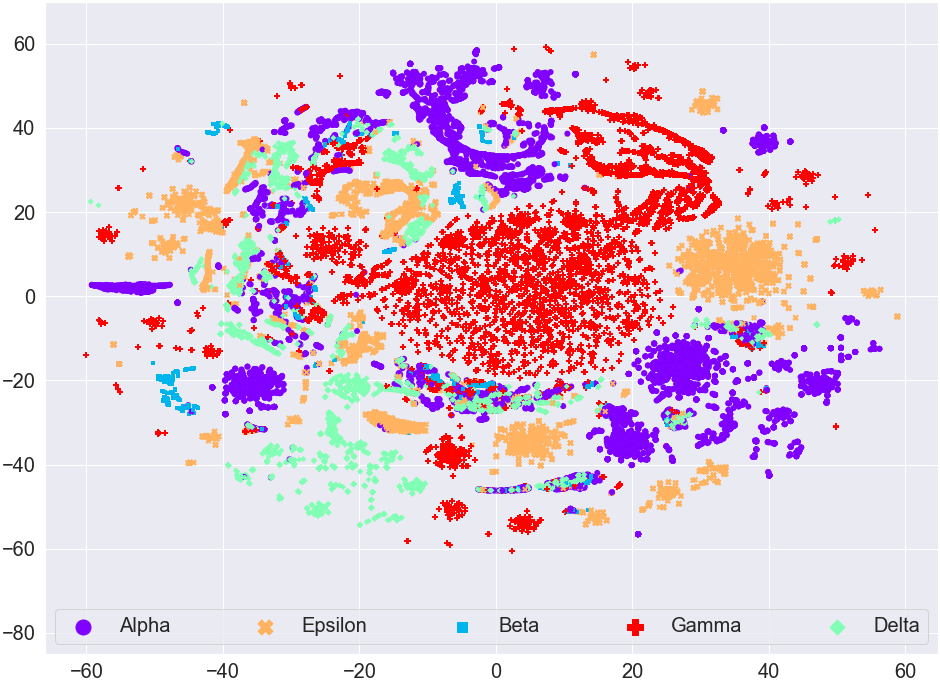}
    \caption{Before Lasso Regression}
  \end{subfigure}%
  \begin{subfigure}{.5\textwidth}
    \centering
    \includegraphics[scale = 0.35]{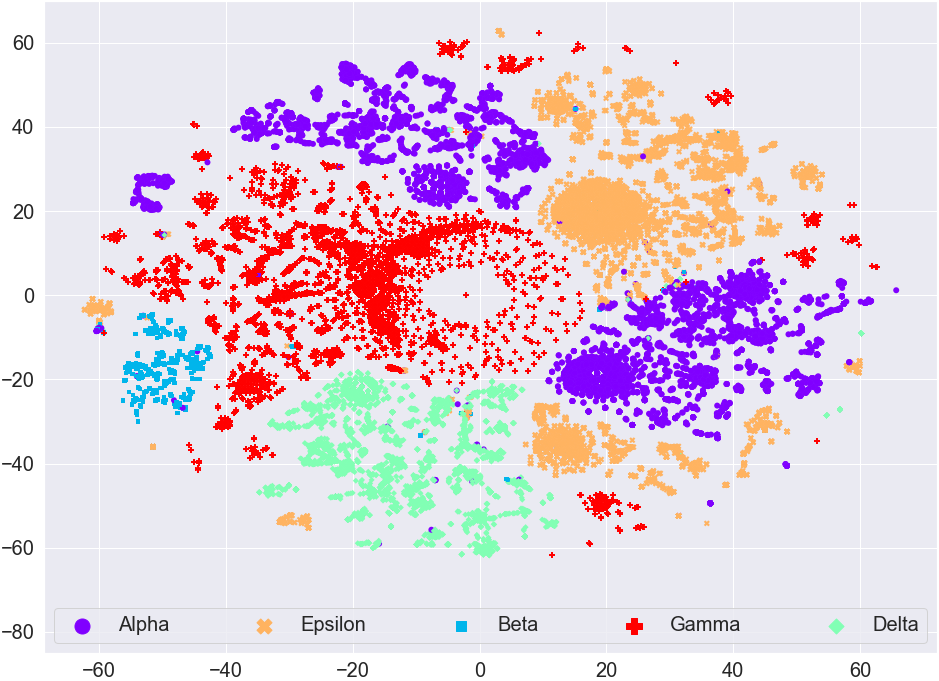}
    \caption{After Lasso Regression}
  \end{subfigure}
  \caption{t-SNE plot for the frequency vectors of GISAID dataset
    along with the variant information.}
  \label{fig_tsne_dataset_plot}
\end{figure}

To see if there is any (hidden) clustering in the data, we mapped the
data to 2D real vectors using the t-SNE approach.  The visualization
using t-SNE for the GISAID dataset (before applying Lasso Regression
for feature selection) is given in Figure~\ref{fig_tsne_dataset_plot}
(a).  We can see that different variants are scattered everywhere.
Although there are small clusters for different variants, we cannot
see a clear (complete) cluster for any variant.  This analysis shows
that any clustering algorithm (like $K$-means) will not work on such
type of data directly.  We need to perform some preprocessing on the
data to cluster the variants effectively.

The visualization using t-SNE for the GISAID dataset (after applying
Lasso Regression for feature selection) is given in
Figure~\ref{fig_tsne_dataset_plot} (b).  We can observe that the
clusters are more clearly visible after applying feature selection as
compared to clustering on frequency vectors without feature selection (as given in
Figure~\ref{fig_tsne_dataset_plot} (a)).  For example, the Epsilon
variant (orange color dots) in Figure~\ref{fig_tsne_dataset_plot}
(a) is scattered everywhere, however, it is concentrated to one place
(top right side) in Figure~\ref{fig_tsne_dataset_plot} (b).  The same
behavior can be observed for the Beta variant (blue color dots).  The
Delta variant (green color) is concentrated on bottom left side in
Figure~\ref{fig_tsne_dataset_plot} (b) forming a single cluster.


\begin{figure}[ht!]
  \centering
  \begin{subfigure}{.5\textwidth}
    \centering
    \includegraphics[scale = 0.35]{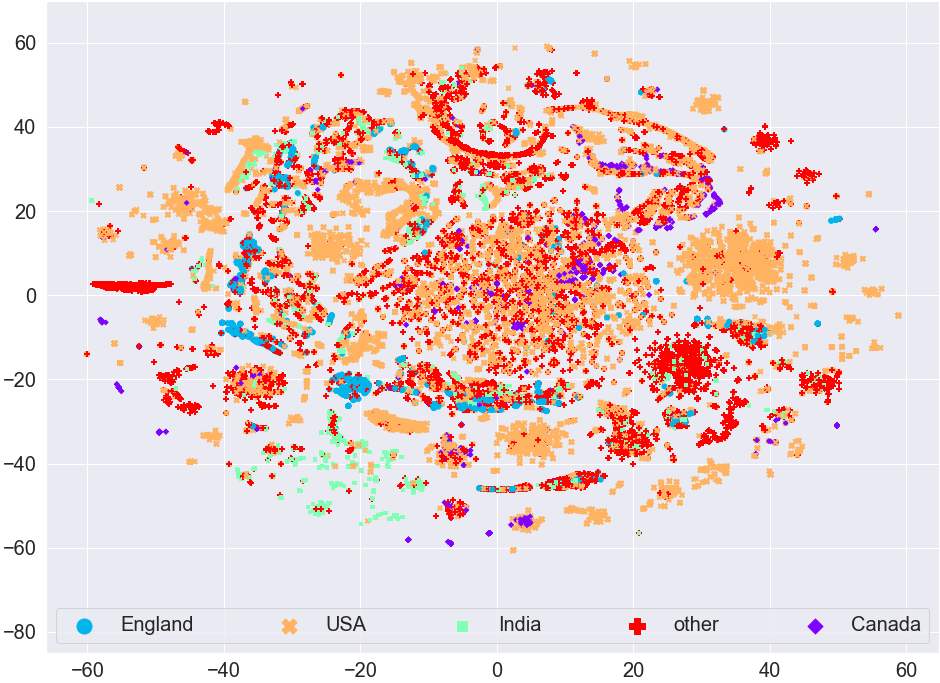}
    \caption{Before Lasso Regression}
  \end{subfigure}%
  \begin{subfigure}{.5\textwidth}
    \centering
    \includegraphics[scale = 0.35]{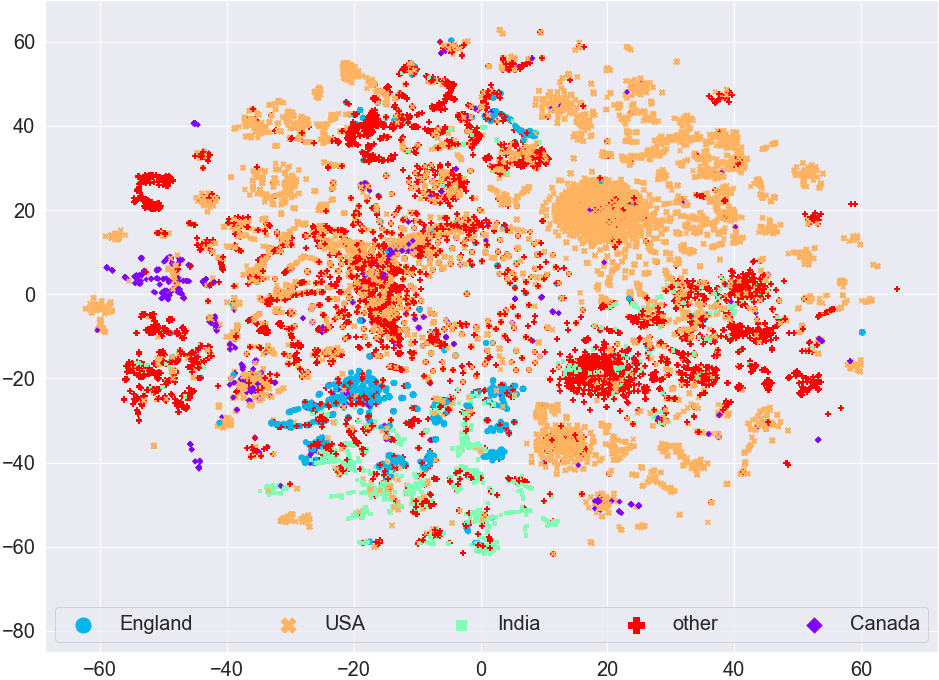}
    \caption{After Lasso Regression}
  \end{subfigure}
  \caption{t-SNE plot for the frequency vectors of GISAID dataset
    along with the country information.}
  \label{fig_country_tsne_dataset_plot}
\end{figure}

We also evaluate if there is any (hidden) clustering in the data
corresponding to different countries.  We took the countries with the
most COVID-19 infected patients and show the t-SNE plots in
Figure~\ref{fig_country_tsne_dataset_plot}.  We can observe that there
is no clear cluster for any country.  Since the mutations in the spike
protein take place for different variants, and a country can have
patients affected by different variants of the coronavirus, we cannot
expect any sort of clustering in this case.  Based on these analyses,
we evaluate the quality of our clusters using variants information
rather than country information.

\begin{figure}[ht!]
  \centering
  \begin{subfigure}{.5\textwidth}
    \centering
    \includegraphics[scale = 0.35]{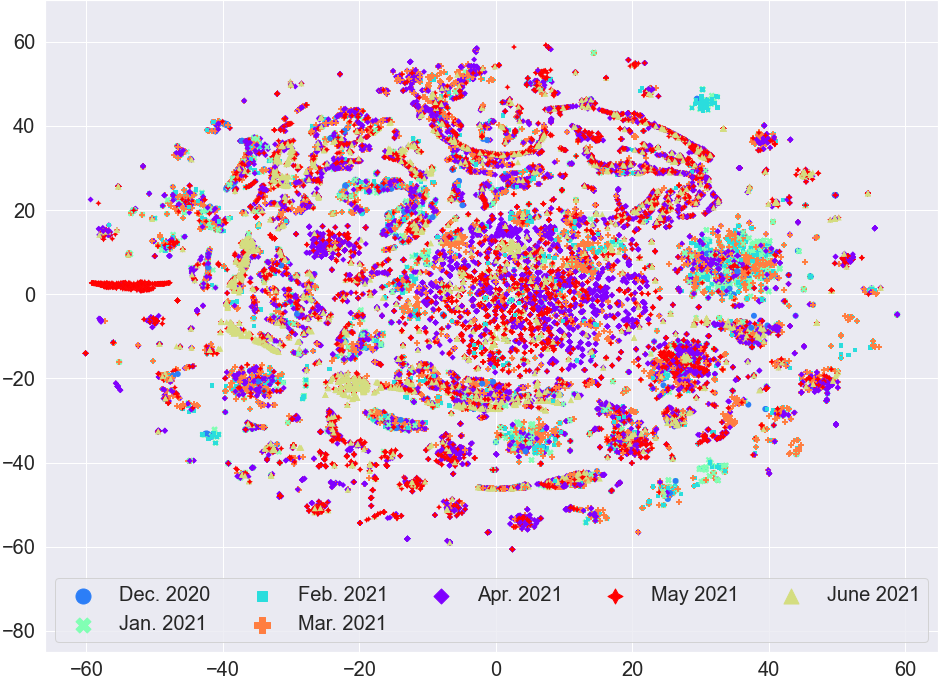}
    \caption{Before Lasso Regression}
  \end{subfigure}%
  \begin{subfigure}{.5\textwidth}
    \centering
    \includegraphics[scale = 0.35]{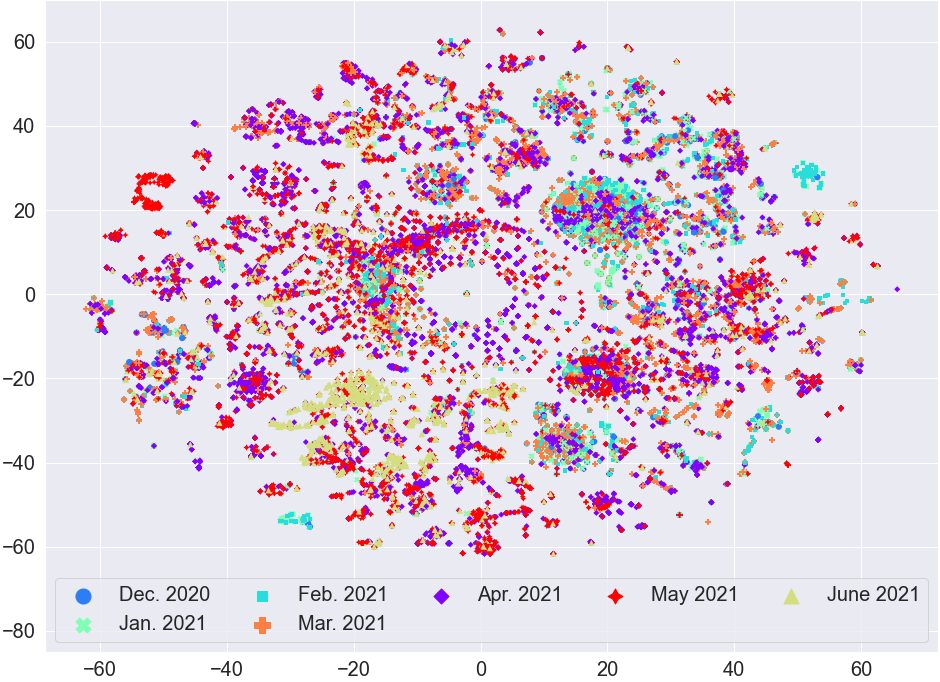}
    \caption{After Lasso Regression}
  \end{subfigure}
  \caption{t-SNE plot for the frequency vectors of GISAID dataset
    along with the months information.}
\label{fig_months_tsne_dataset_plot}
\end{figure}

The t-SNE plots for feature vectors (with and without feature
selection using Lasso Regression) with month information are given in
Figure~\ref{fig_months_tsne_dataset_plot}.  We can again observe that
there is no clear clustering for any month in the data (even after the
feature selection method).  Therefore, since we are unable to find any
useful pattern in the data with respect to month information, we can
say that the spread of coronavirus has nothing to do with different
seasons over the years (e.g., summer, spring, winter, etc.).

\section{Results and Discussion}
\label{sec_results_and_discussion}

To compute the quality of a clustering, we use the (weighted) F1
score.  We label each cluster using the variant which labels the
majority of sequences in the cluster (e.g., if the majority of the
sequences in a cluster belongs to Alpha variant, we assign the label
`Alpha' to that cluster).  Now, using these assigned labels, we
compute the F1 score for each variant separately.
The (weighted) F1 scores for different methods are given in
Table~\ref{tbl_f1_weighted}.  We can see that the F1 scores for
$K$-means + Ridge are better than simple $K$-means (the baseline),
however, they are not better than $K$-means + Lasso for the majority
of the variants.  One behavior that we can observe for all methods is
a low F1 score for the Beta variant.  This is due to the much smaller
number of sequences available for the Beta variant (see
Table~\ref{tbl_variant_information}).
 
\begin{table}[ht!]
  \centering
  \begin{tabular}{lccccc}
    \hline
    & \multicolumn{5}{c}{F1 Score (Weighted) for Different Variants} \\
    \cline{2-6}
    Methods & Alpha & Beta & Delta &  Gamma & Epsilon \\
    \hline	\hline	
    K-means & 0.3598 & 0.1070 & 0.6110 & 0.6908 & 0.6527 \\
    K-means + Ridge& 0.9992 & 0.0058 & 0.8643 & 0.9998 & 0.7748 \\
    K-means + Lasso & 0.9987 & 0.2705 & 0.9991 & 0.9998 & 0.9704 \\
    \hline
  \end{tabular}
  \caption{Variant-wise F1 (weighted) score for different clustering
    methods with $k = 5$ for K-means.}
  \label{tbl_f1_weighted}
\end{table}

The contingency tables of variants versus clusters for $K$-means only,
$K$-means + Ridge Regression, and $K$-means + Lasso Regression are
given in Table~\ref{tbl_contingency_kmeans_only}.  We can see that the
variants are not clearly clustered into separate groups if we only use
$K$-means without any feature selection method.  However, with the
feature selection methods (Ridge and Lasso Regression), we can clearly
observe that different variants are grouped into their respective
clusters.  We can also observe that Lasso Regression based feature
selection gives us more accurate clusters than Ridge Regression based
feature selection.  Although Lasso Regression gave us $964$ features
out of $4977$ as compared to $1242$ features given by Ridge
Regression, those $964$ features are a more accurate representation of
the original data (as observed in
Table~\ref{tbl_contingency_kmeans_only}).

\begin{table}[ht!]
  \centering
  \begin{tabular}{p{1cm}p{0.4cm}p{0.6cm}p{0.4cm}p{0.4cm}p{0.6cm}|p{0.6cm}p{0.6cm}p{0.6cm}p{0.6cm}p{0.6cm}|ccccc}
    \hline
    & \multicolumn{5}{c}{K-means (Clust. IDs)} & \multicolumn{5}{c}{K-means + Ridge (Clust. IDs)} & \multicolumn{5}{c}{K-means + Lasso (Clust. IDs)} \\
    \cline{2-6} \cline{7-11} \cline{12-16}
      Variants & 0 & 1 & 2 & 3 & 4 & 0 & 1 & 2 & 3 & 4 & 0 & 1 & 2 & 3 & 4 \\
      \hline	\hline
        Alpha & 1512 & 8762 & 86 & 680 & 2926 & 6 & 11415 & 310 & 344 & 1891 & 6 & 74 & 13622 & 258 & 6   \\
        Beta & 295 & 601 & 33 & 172 & 626 & 1 & 4 & 1596 & 109 & 17 & 4 & 37 & 10 & 1673 & 3  \\
        Epsilon & 956 & 7848 & 187 & 638 & 3155 & 0 & 1 & 8688 & 654 & 3441 & 0 & 4076 & 1 & 8705 & 2 \\
        Delta & 2706 & 2605 & 30 & 868 & 1342 & 0 & 0 & 3126 & 3996 & 429 & 0 & 111 & 0 & 45 & 7395 \\
        Gamma & 682 & 22140 & 50 & 741 & 3016 & 26426 & 13 & 16 & 147 & 27 & 26566 & 26 & 24 & 12 & 1 \\
      \hline
  \end{tabular}
  \caption{Contingency tables of variants vs clusters.}
  \label{tbl_contingency_kmeans_only}
\end{table}

\subsection{Optimal value of K for K-means}\label{sec_optimal_num_of_clust}

\begin{figure}[ht!]
    \centering
    \includegraphics[scale = 0.5]{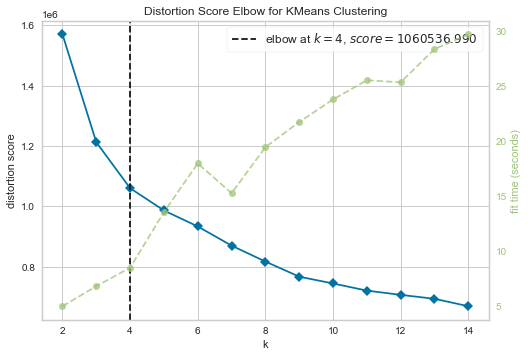}
    \caption{Sum of squared distances (distortion score) for different
      numbers of clusters using $K$-means.  The blue line shows the
      distortion score, while the dashed green line shows the runtime
      (in sec.) for different number of clusters (on the x-axis).  The
      dashed black line shows the optimal number of clusters computed
      using Elbow method~\cite{satopaa2011finding}.}
    \label{fig_ideal_k_value}
\end{figure}

We use the elbow method to determine the optimal number of clusters
(value of $K$ for $K$-means) by fitting the model with values of $K$
ranging from $2$ to $14$.  The quality measure that we use for this
analysis is `\textit{distortion}' that computes the sum of squared
distances from each point to its assigned center.
Figure~\ref{fig_ideal_k_value} shows the distortion score for
different values of $K$.  To observe the trade-off between the runtime
and distortion score, we also plot the (fit/training) runtime (in
seconds) in Figure~\ref{fig_ideal_k_value} (dashed green line).  To
determine the optimal value for $K$, we use ``knee point detection
algorithm (KPDA)"~\cite{satopaa2011finding}.  We can see in
Figure~\ref{fig_ideal_k_value} that the ideal number of clusters is
$4$.

Since we know that we have $5$ variants in our data, we take $K = 5$
throughout the experiments.  It is good, however, to have this
independent way of discovering the optimal $K$, and that it is close
to the true number.  The most likely reason why KPDA selected $4$ as
the ideal number of clusters is because the Beta variant is not well
represented in our data (see Table~\ref{tbl_variant_information}).
Moreover, the $K$-means algorithm is not very successful in clustering
the Beta variant (as can be seen from its F1 score in
Table~\ref{tbl_f1_weighted}).

\subsection{The Spread of Variants Over Time}

\begin{figure}[ht!]
  \centering
  \includegraphics[scale = 0.6]{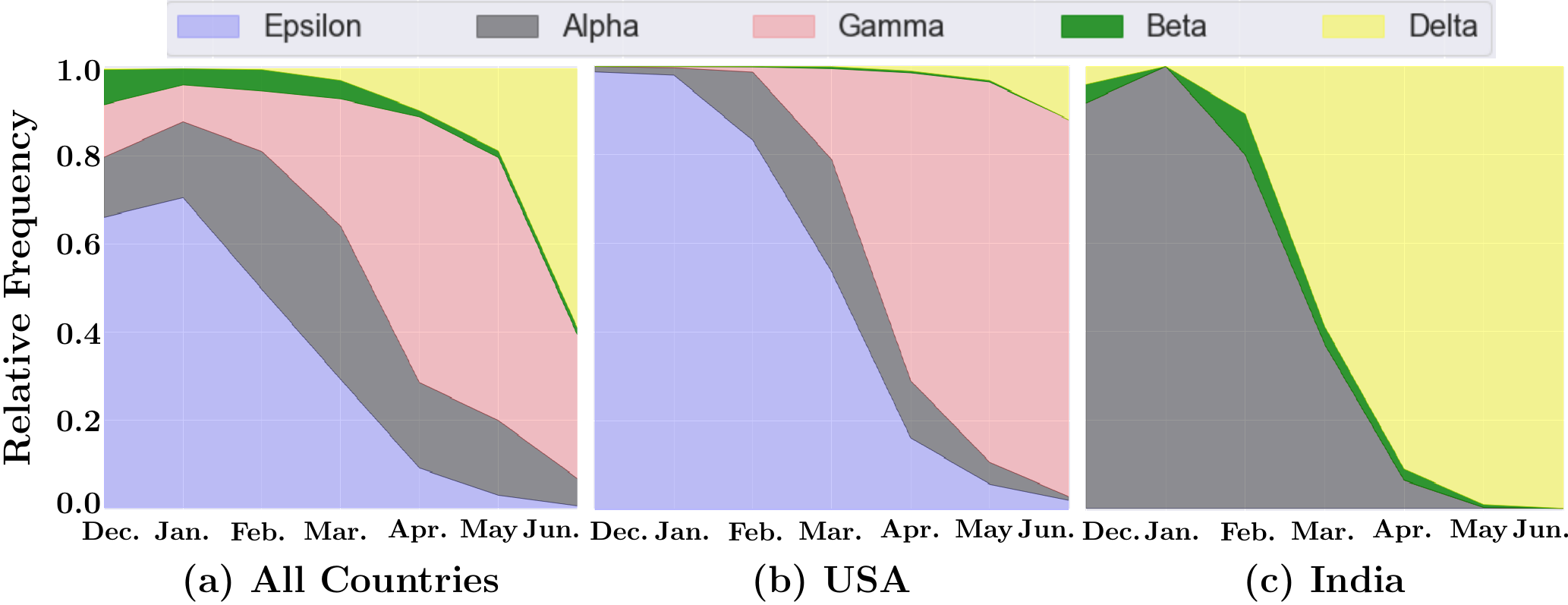}
  \caption{Month-wise relative frequency of each variant. We observed the relative frequencies of variants for different months separately (from December 2020 to June 2021).}
    \label{fig_relative_freq_lasso}
\end{figure}

In this section, we analyze the behavior of each variant by
considering time as the changing factor.  More precisely, we show how
the rate of spread of each variant is changing as time passes (from
December 2020 to June 2021).  Figure~\ref{fig_relative_freq_lasso} (a)
shows the relative frequencies of different variants corresponding to
different months (for all countries in the data).  
This study helps us analyze which variants are increasing in number and
which ones are waning over time.  We can observe that initially, the
Delta variant was very rare (in December 2019).  However, in the
middle of the year 2021 (from April to June), we can see that the
Delta variant is spreading very quickly.  The behavior of the Epsilon
variant is opposite to the Delta variant.
We can also observe that the Beta variant is not spreading at an
alarming rate, hence there is no threat of the spread of that variant
in the near future.

Figure~\ref{fig_relative_freq_lasso} (b) and
Figure~\ref{fig_relative_freq_lasso} (c) shows the relative frequency
of different variants in USA and India, respectively, from December
2020 to June 2021.  We can observe that the Epsilon variant
(California variant) was initially present in high numbers in the USA.
However, as time passes (and the vaccination process speeds up), we
can see the drop in the Epsilon variant.  We can also observe that
from February 2021 onward, the Gamma variant is spreading at an
alarming rate.
In the case of India, we can see that, initially, the Alpha variant
was the variant of concern.  However, after January 2021, Delta
becomes the dominantly spreading variant.

\subsection{Importance of Each Amino Acid}

\begin{figure}[ht!]
  \centering
  \centering
  \includegraphics{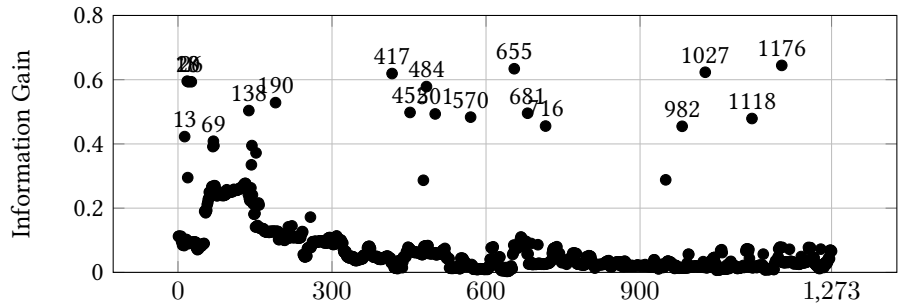}
  \caption{IG for amino acids with respect to
    variants. $x$-axis corresponds to amino acid positions in the
    spike sequences.}
  \label{fig_data_correlation}
\end{figure}

We compute Information Gain (IG) between each attribute (amino acid
position) and the class (variant).  The IG is defined as
$IG(Class,position) = H(Class) - H(Class | position)$, where $H=
\sum_{ i \in Class} -p_i \log p_i$ is the entropy, and $p_i$ is the
probability of the class $i$.
The IG values for each attribute are shown in
Figure~\ref{fig_data_correlation} (a higher value is better).  The
USA's Centers for Disease Control and Prevention (CDC) declared
mutations at certain positions from one variant to the
other~\cite{CDS_variantDef}.  We use their mutation information to
compare them with the attributes having high IG values in
Figure~\ref{fig_data_correlation}.  We note that the our high IG value
attributes are the same as given by CDC.  For example, R452L is
present in Epsilon, Delta, and Kappa lineages and sub-lineages.
Similarly, K417N, E484K, and N501Y substitutions are present in Beta
variant while K417T, E484K, and N501Y substitutions are present in
Gamma variant~\cite{CDS_variantDef} (see
Figure~\ref{fig_data_correlation}).  The IG values for each attribute
are available online for further
analysis\footnote{\url{https://github.com/sarwanpasha/Visualization_covid_data/blob/main/correlation_data.csv}}.

\section{Conclusion}
\label{sec_conclusion}
We propose a method to effectively cluster the
variants of SARS-CoV-2 using amino acid sequences corresponding to the
spike region.  Our results show that efficient feature selection
methods such as Ridge Regression and Lasso Regression greatly help
improve the performance of $K$-means clustering.  We also show the
rate of spread of each variant with time in different
countries. Finally, we show the importance of each amino acid position
by computing information gain and compare them with the CDC's provided
information for these same positions.
In the future, we will work towards using more SARS-CoV-2 sequences
with more variants, as the data continues to accumulate.  Using un-supervised feature selection methods would be another
interesting future work.

\bibliographystyle{ACM-Reference-Format}
\bibliography{COVID19Visualization}


\begin{thebibliography}{47}


\ifx \showCODEN    \undefined \def \showCODEN     #1{\unskip}     \fi
\ifx \showDOI      \undefined \def \showDOI       #1{#1}\fi
\ifx \showISBNx    \undefined \def \showISBNx     #1{\unskip}     \fi
\ifx \showISBNxiii \undefined \def \showISBNxiii  #1{\unskip}     \fi
\ifx \showISSN     \undefined \def \showISSN      #1{\unskip}     \fi
\ifx \showLCCN     \undefined \def \showLCCN      #1{\unskip}     \fi
\ifx \shownote     \undefined \def \shownote      #1{#1}          \fi
\ifx \showarticletitle \undefined \def \showarticletitle #1{#1}   \fi
\ifx \showURL      \undefined \def \showURL       {\relax}        \fi
\providecommand\bibfield[2]{#2}
\providecommand\bibinfo[2]{#2}
\providecommand\natexlab[1]{#1}
\providecommand\showeprint[2][]{arXiv:#2}

\bibitem[\protect\citeauthoryear{Ahmad, Ali, Tariq, Khan, Shabbir, and
  Zaman}{Ahmad et~al\mbox{.}}{2020}]%
        {AHMAD2020Combinatorial}
\bibfield{author}{\bibinfo{person}{M. Ahmad}, \bibinfo{person}{S. Ali},
  \bibinfo{person}{J. Tariq}, \bibinfo{person}{I. Khan}, \bibinfo{person}{M.
  Shabbir}, {and} \bibinfo{person}{A. Zaman}.} \bibinfo{year}{2020}\natexlab{}.
\newblock \showarticletitle{Combinatorial trace method for network
  immunization}.
\newblock \bibinfo{journal}{\emph{Information Sciences}}  \bibinfo{volume}{519}
  (\bibinfo{year}{2020}), \bibinfo{pages}{215 -- 228}.
\newblock


\bibitem[\protect\citeauthoryear{Ahmad, Tariq, Farhan, Shabbir, and Khan}{Ahmad
  et~al\mbox{.}}{2016}]%
        {Ahmad2016AusDM}
\bibfield{author}{\bibinfo{person}{M. Ahmad}, \bibinfo{person}{J. Tariq},
  \bibinfo{person}{M. Farhan}, \bibinfo{person}{M. Shabbir}, {and}
  \bibinfo{person}{I. Khan}.} \bibinfo{year}{2016}\natexlab{}.
\newblock \showarticletitle{Who should receive the vaccine?}. In
  \bibinfo{booktitle}{\emph{Australasian Data Mining Conference (AusDM)}}.
  \bibinfo{pages}{137--145}.
\newblock


\bibitem[\protect\citeauthoryear{Ahmad, Tariq, Shabbir, and Khan}{Ahmad
  et~al\mbox{.}}{2017}]%
        {ahmad2017spectral}
\bibfield{author}{\bibinfo{person}{M. Ahmad}, \bibinfo{person}{J. Tariq},
  \bibinfo{person}{M. Shabbir}, {and} \bibinfo{person}{I. Khan}.}
  \bibinfo{year}{2017}\natexlab{}.
\newblock \showarticletitle{Spectral Methods for Immunization of Large
  Networks}.
\newblock \bibinfo{journal}{\emph{Australasian Journal of Information Systems}}
   \bibinfo{volume}{21} (\bibinfo{year}{2017}).
\newblock


\bibitem[\protect\citeauthoryear{Ali}{Ali}{2021}]%
        {ali2021cache}
\bibfield{author}{\bibinfo{person}{S. Ali}.} \bibinfo{year}{2021}\natexlab{}.
\newblock \showarticletitle{Cache Replacement Algorithm}.
\newblock \bibinfo{journal}{\emph{arXiv preprint arXiv:2107.14646}}
  (\bibinfo{year}{2021}).
\newblock


\bibitem[\protect\citeauthoryear{Ali, Alvi, Faizullah, Khan, Alshanqiti, and
  Khan}{Ali et~al\mbox{.}}{2020a}]%
        {Ali2019Detecting1}
\bibfield{author}{\bibinfo{person}{S. Ali}, \bibinfo{person}{M. Alvi},
  \bibinfo{person}{S. Faizullah}, \bibinfo{person}{M. Khan},
  \bibinfo{person}{A. Alshanqiti}, {and} \bibinfo{person}{I. Khan}.}
  \bibinfo{year}{2020}\natexlab{a}.
\newblock \showarticletitle{Detecting DDoS Attack on SDN Due to Vulnerabilities
  in OpenFlow}. In \bibinfo{booktitle}{\emph{International Conference on
  Advances in the Emerging Computing Technologies (AECT)}}.
  \bibinfo{pages}{1--6}.
\newblock


\bibitem[\protect\citeauthoryear{Ali, Ciccolella, Lucarella, D.~Vedova, and
  Patterson}{Ali et~al\mbox{.}}{2021a}]%
        {ali2021simpler}
\bibfield{author}{\bibinfo{person}{S. Ali}, \bibinfo{person}{S. Ciccolella},
  \bibinfo{person}{L. Lucarella}, \bibinfo{person}{G. D.~Vedova}, {and}
  \bibinfo{person}{M.~D. Patterson}.} \bibinfo{year}{2021}\natexlab{a}.
\newblock \showarticletitle{Simpler and Faster Development of Tumor Phylogeny
  Pipelines}.
\newblock \bibinfo{journal}{\emph{bioRxiv}} (\bibinfo{year}{2021}).
\newblock


\bibitem[\protect\citeauthoryear{Ali, Mansoor, Arshad, and Khan}{Ali
  et~al\mbox{.}}{2019}]%
        {ali2019short}
\bibfield{author}{\bibinfo{person}{S. Ali}, \bibinfo{person}{H. Mansoor},
  \bibinfo{person}{N. Arshad}, {and} \bibinfo{person}{I. Khan}.}
  \bibinfo{year}{2019}\natexlab{}.
\newblock \showarticletitle{Short term load forecasting using smart meter
  data}. In \bibinfo{booktitle}{\emph{International Conference on Future Energy
  Systems (e-Energy)}}. \bibinfo{pages}{419--421}.
\newblock


\bibitem[\protect\citeauthoryear{Ali, Mansoor, Khan, Arshad, Khan, and
  Faizullah}{Ali et~al\mbox{.}}{2020b}]%
        {Ali2020ShortTerm}
\bibfield{author}{\bibinfo{person}{S. Ali}, \bibinfo{person}{H. Mansoor},
  \bibinfo{person}{I. Khan}, \bibinfo{person}{N. Arshad}, \bibinfo{person}{M.
  Khan}, {and} \bibinfo{person}{S. Faizullah}.}
  \bibinfo{year}{2020}\natexlab{b}.
\newblock \showarticletitle{Short-Term Load Forecasting Using AMI Data}.
\newblock \bibinfo{journal}{\emph{CoRR}}  \bibinfo{volume}{abs/1912.12479}
  (\bibinfo{year}{2020}).
\newblock


\bibitem[\protect\citeauthoryear{Ali, Sahoo, Ullah, Zelikovskiy, Patterson, and
  Khan}{Ali et~al\mbox{.}}{2021b}]%
        {ali2021k}
\bibfield{author}{\bibinfo{person}{S. Ali}, \bibinfo{person}{B. Sahoo},
  \bibinfo{person}{N. Ullah}, \bibinfo{person}{A. Zelikovskiy},
  \bibinfo{person}{M.~D. Patterson}, {and} \bibinfo{person}{I. Khan}.}
  \bibinfo{year}{2021}\natexlab{b}.
\newblock \showarticletitle{A k-mer Based Approach for SARS-CoV-2 Variant
  Identification}.
\newblock \bibinfo{journal}{\emph{arXiv preprint arXiv:2108.03465}}
  (\bibinfo{year}{2021}).
\newblock


\bibitem[\protect\citeauthoryear{Ali, Shakeel, Khan, Faizullah, and Khan}{Ali
  et~al\mbox{.}}{2021c}]%
        {ali2021predicting}
\bibfield{author}{\bibinfo{person}{S. Ali}, \bibinfo{person}{M. Shakeel},
  \bibinfo{person}{I. Khan}, \bibinfo{person}{S. Faizullah}, {and}
  \bibinfo{person}{M. Khan}.} \bibinfo{year}{2021}\natexlab{c}.
\newblock \showarticletitle{Predicting attributes of nodes using network
  structure}.
\newblock \bibinfo{journal}{\emph{ACM Transactions on Intelligent Systems and
  Technology (TIST)}} \bibinfo{volume}{12}, \bibinfo{number}{2}
  (\bibinfo{year}{2021}), \bibinfo{pages}{1--23}.
\newblock


\bibitem[\protect\citeauthoryear{Atzori et~al\mbox{.}}{Atzori
  et~al\mbox{.}}{2014}]%
        {atzori2014electromyography}
\bibfield{author}{\bibinfo{person}{M. Atzori} {et~al\mbox{.}}}
  \bibinfo{year}{2014}\natexlab{}.
\newblock \showarticletitle{Electromyography data for non-invasive
  naturally-controlled robotic hand prostheses}.
\newblock \bibinfo{journal}{\emph{Sci. data}} \bibinfo{volume}{1},
  \bibinfo{number}{1} (\bibinfo{year}{2014}), \bibinfo{pages}{1--13}.
\newblock


\bibitem[\protect\citeauthoryear{Blaisdell}{Blaisdell}{1986}]%
        {Blaisdell1986AMeasureOfSimilarity}
\bibfield{author}{\bibinfo{person}{B. Blaisdell}.}
  \bibinfo{year}{1986}\natexlab{}.
\newblock \showarticletitle{A measure of the similarity of sets of sequences
  not requiring sequence alignment}.
\newblock \bibinfo{journal}{\emph{Proceedings of the National Academy of
  Sciences}}  \bibinfo{volume}{83} (\bibinfo{year}{1986}),
  \bibinfo{pages}{5155--5159}.
\newblock


\bibitem[\protect\citeauthoryear{Chang et~al\mbox{.}}{Chang
  et~al\mbox{.}}{2014}]%
        {Chang2014ANA}
\bibfield{author}{\bibinfo{person}{G. Chang} {et~al\mbox{.}}}
  \bibinfo{year}{2014}\natexlab{}.
\newblock \showarticletitle{A novel alignment-free method for genome
  analysis:HIV-1 subtyping and HEV genotyping}.
\newblock \bibinfo{journal}{\emph{Information Sciences}}  \bibinfo{volume}{279}
  (\bibinfo{year}{2014}), \bibinfo{pages}{776--784}.
\newblock


\bibitem[\protect\citeauthoryear{Chowdhury and Garai}{Chowdhury and
  Garai}{2017}]%
        {Chowdhury2017MultipleSequences}
\bibfield{author}{\bibinfo{person}{B. Chowdhury} {and} \bibinfo{person}{G.
  Garai}.} \bibinfo{year}{2017}\natexlab{}.
\newblock \showarticletitle{A review on multiple sequence alignment from the
  perspective of genetic algorithm}.
\newblock \bibinfo{journal}{\emph{Genomics}}  \bibinfo{volume}{1}
  (\bibinfo{year}{2017}), \bibinfo{pages}{419--431}.
\newblock


\bibitem[\protect\citeauthoryear{{CNN Health, Tracking Covid-19 global
  spread}}{{CNN Health, Tracking Covid-19 global spread}}{2021}]%
        {covid_statistics_cnn}
\bibfield{author}{\bibinfo{person}{{CNN Health, Tracking Covid-19 global
  spread}}.} \bibinfo{year}{2021}\natexlab{}.
\newblock
  \bibinfo{howpublished}{\url{https://edition.cnn.com/interactive/2020/health/coronavirus-maps-and-cases/}}.
\newblock
\newblock
\shownote{[Online; accessed 4-September-2021].}


\bibitem[\protect\citeauthoryear{Devijver and Kittler}{Devijver and
  Kittler}{1982}]%
        {validationSetApproach}
\bibfield{author}{\bibinfo{person}{P. Devijver} {and} \bibinfo{person}{J.
  Kittler}.} \bibinfo{year}{1982}\natexlab{}.
\newblock \showarticletitle{Pattern Recognition: A Statistical Approach}. In
  \bibinfo{booktitle}{\emph{London, GB: Prentice-Hall}}.
  \bibinfo{pages}{1--448}.
\newblock


\bibitem[\protect\citeauthoryear{du~Plessis et~al\mbox{.}}{du~Plessis
  et~al\mbox{.}}{2021}]%
        {du2021establishment}
\bibfield{author}{\bibinfo{person}{L. du Plessis} {et~al\mbox{.}}}
  \bibinfo{year}{2021}\natexlab{}.
\newblock \showarticletitle{Establishment and lineage dynamics of the
  SARS-CoV-2 epidemic in the UK}.
\newblock \bibinfo{journal}{\emph{Science}} \bibinfo{volume}{371},
  \bibinfo{number}{6530} (\bibinfo{year}{2021}), \bibinfo{pages}{708--712}.
\newblock


\bibitem[\protect\citeauthoryear{Dwivedi and Sengupta}{Dwivedi and
  Sengupta}{2012}]%
        {Dwivedi2012ClassificationOH}
\bibfield{author}{\bibinfo{person}{S.~K. Dwivedi} {and} \bibinfo{person}{S.
  Sengupta}.} \bibinfo{year}{2012}\natexlab{}.
\newblock \showarticletitle{Classification of HIV-1 Sequences Using Profile
  Hidden Markov Models}.
\newblock \bibinfo{journal}{\emph{PLoS ONE}}  \bibinfo{volume}{7}
  (\bibinfo{year}{2012}).
\newblock


\bibitem[\protect\citeauthoryear{Farhan, Tariq, Zaman, Shabbir, and
  Khan}{Farhan et~al\mbox{.}}{2017}]%
        {farhan2017efficient}
\bibfield{author}{\bibinfo{person}{M. Farhan}, \bibinfo{person}{J. Tariq},
  \bibinfo{person}{A. Zaman}, \bibinfo{person}{M. Shabbir}, {and}
  \bibinfo{person}{I. Khan}.} \bibinfo{year}{2017}\natexlab{}.
\newblock \showarticletitle{Efficient Approximation Algorithms for Strings
  Kernel Based Sequence Classification}. In \bibinfo{booktitle}{\emph{Advances
  in neural information processing systems (NeurIPS)}}. \bibinfo{publisher}{.},
  \bibinfo{pages}{6935--6945}.
\newblock


\bibitem[\protect\citeauthoryear{Galloway et~al\mbox{.}}{Galloway
  et~al\mbox{.}}{2021}]%
        {galloway2021emergence}
\bibfield{author}{\bibinfo{person}{S. Galloway} {et~al\mbox{.}}}
  \bibinfo{year}{2021}\natexlab{}.
\newblock \showarticletitle{Emergence of SARS-CoV-2 b. 1.1. 7 lineage}.
\newblock \bibinfo{journal}{\emph{Morbidity and Mortality Weekly Report}}
  \bibinfo{volume}{70}, \bibinfo{number}{3} (\bibinfo{year}{2021}),
  \bibinfo{pages}{95}.
\newblock


\bibitem[\protect\citeauthoryear{{GISAID Website}}{{GISAID Website}}{2021}]%
        {gisaid_website_url}
\bibfield{author}{\bibinfo{person}{{GISAID Website}}.}
  \bibinfo{year}{2021}\natexlab{}.
\newblock \bibinfo{howpublished}{\url{https://www.gisaid.org/}}.
\newblock
\newblock
\shownote{[Online; accessed 4-September-2021].}


\bibitem[\protect\citeauthoryear{Hadfield, Megill, Bell, Huddleston, Potter,
  Callender, Sagulenko, Bedford, and Neher}{Hadfield et~al\mbox{.}}{2018}]%
        {hadfield2018a}
\bibfield{author}{\bibinfo{person}{J. Hadfield}, \bibinfo{person}{C. Megill},
  \bibinfo{person}{S.M. Bell}, \bibinfo{person}{J. Huddleston},
  \bibinfo{person}{B. Potter}, \bibinfo{person}{C. Callender},
  \bibinfo{person}{P. Sagulenko}, \bibinfo{person}{T. Bedford}, {and}
  \bibinfo{person}{R.A. Neher}.} \bibinfo{year}{2018}\natexlab{}.
\newblock \showarticletitle{Nextstrain: real-time tracking of pathogen
  evolution}.
\newblock \bibinfo{journal}{\emph{Bioinformatics}}  \bibinfo{volume}{34}
  (\bibinfo{year}{2018}), \bibinfo{pages}{4121--4123}.
\newblock


\bibitem[\protect\citeauthoryear{Hassan, Khan, Shabbir, and Abbas}{Hassan
  et~al\mbox{.}}{2021}]%
        {Hassan2021Computing}
\bibfield{author}{\bibinfo{person}{Z. Hassan}, \bibinfo{person}{I. Khan},
  \bibinfo{person}{M. Shabbir}, {and} \bibinfo{person}{W. Abbas}.}
  \bibinfo{year}{2021}\natexlab{}.
\newblock \bibinfo{title}{Computing Graph Descriptors on Edge Streams}.
  (\bibinfo{year}{2021}).
\newblock
\newblock
\shownote{\url{https://www.researchgate.net/publication/353671195_Computing_Graph_Descriptors_on_Edge_Streams}.}


\bibitem[\protect\citeauthoryear{Hassan, Shabbir, Khan, and Abbas}{Hassan
  et~al\mbox{.}}{2020}]%
        {hassan2020estimating}
\bibfield{author}{\bibinfo{person}{Z.R Hassan}, \bibinfo{person}{M. Shabbir},
  \bibinfo{person}{I. Khan}, {and} \bibinfo{person}{W. Abbas}.}
  \bibinfo{year}{2020}\natexlab{}.
\newblock \showarticletitle{Estimating Descriptors for Large Graphs}. In
  \bibinfo{booktitle}{\emph{Advances in Knowledge Discovery and Data Mining
  (PAKDD)}}. \bibinfo{pages}{779--791}.
\newblock


\bibitem[\protect\citeauthoryear{Hoerl, Kannard, and Baldwin}{Hoerl
  et~al\mbox{.}}{1975}]%
        {hoerl1975ridge}
\bibfield{author}{\bibinfo{person}{Arthur~E Hoerl}, \bibinfo{person}{Robert~W
  Kannard}, {and} \bibinfo{person}{Kent~F Baldwin}.}
  \bibinfo{year}{1975}\natexlab{}.
\newblock \showarticletitle{Ridge regression: some simulations}.
\newblock \bibinfo{journal}{\emph{Communications in Statistics-Theory and
  Methods}} \bibinfo{volume}{4}, \bibinfo{number}{2} (\bibinfo{year}{1975}),
  \bibinfo{pages}{105--123}.
\newblock


\bibitem[\protect\citeauthoryear{Krishnan, Kamath, and Sugumaran}{Krishnan
  et~al\mbox{.}}{2021}]%
        {Krishnan2021PredictingVaccineHesitancy}
\bibfield{author}{\bibinfo{person}{G. Krishnan}, \bibinfo{person}{S. Kamath},
  {and} \bibinfo{person}{V. Sugumaran}.} \bibinfo{year}{2021}\natexlab{}.
\newblock \showarticletitle{Predicting Vaccine Hesitancy and Vaccine Sentiment
  Using Topic Modeling and Evolutionary Optimization}. In
  \bibinfo{booktitle}{\emph{International Conference on Applications of Natural
  Language to Information Systems (NLDB)}}. \bibinfo{pages}{255--263}.
\newblock


\bibitem[\protect\citeauthoryear{Kuksa, Khan, and Pavlovic}{Kuksa
  et~al\mbox{.}}{2012}]%
        {Kuksa_SequenceKernel}
\bibfield{author}{\bibinfo{person}{P. Kuksa}, \bibinfo{person}{I. Khan}, {and}
  \bibinfo{person}{V. Pavlovic}.} \bibinfo{year}{2012}\natexlab{}.
\newblock \showarticletitle{Generalized Similarity Kernels for Efficient
  Sequence Classification}. In \bibinfo{booktitle}{\emph{SIAM International
  Conference on Data Mining (SDM)}}. \bibinfo{pages}{873--882}.
\newblock


\bibitem[\protect\citeauthoryear{Kuzmin et~al\mbox{.}}{Kuzmin
  et~al\mbox{.}}{2020}]%
        {kuzmin2020machine}
\bibfield{author}{\bibinfo{person}{K. Kuzmin} {et~al\mbox{.}}}
  \bibinfo{year}{2020}\natexlab{}.
\newblock \showarticletitle{Machine learning methods accurately predict host
  specificity of coronaviruses based on spike sequences alone}.
\newblock \bibinfo{journal}{\emph{Biochemical and Biophysical Research
  Communications}}  \bibinfo{volume}{533} (\bibinfo{year}{2020}),
  \bibinfo{pages}{553--558}.
\newblock


\bibitem[\protect\citeauthoryear{Leslie, Eskin, Weston, and Noble}{Leslie
  et~al\mbox{.}}{2003}]%
        {leslie2002mismatch}
\bibfield{author}{\bibinfo{person}{C. Leslie}, \bibinfo{person}{E. Eskin},
  \bibinfo{person}{J. Weston}, {and} \bibinfo{person}{W. Noble}.}
  \bibinfo{year}{2003}\natexlab{}.
\newblock \showarticletitle{Mismatch string kernels for SVM protein
  classification}. In \bibinfo{booktitle}{\emph{Advances in neural information
  processing systems (NeurIPS)}}. \bibinfo{pages}{1441--1448}.
\newblock


\bibitem[\protect\citeauthoryear{Melnyk et~al\mbox{.}}{Melnyk
  et~al\mbox{.}}{2020}]%
        {melnyk2020clustering}
\bibfield{author}{\bibinfo{person}{A. Melnyk} {et~al\mbox{.}}}
  \bibinfo{year}{2020}\natexlab{}.
\newblock \showarticletitle{Clustering based identification of SARS-CoV-2
  subtypes}. In \bibinfo{booktitle}{\emph{International Conference on
  Computational Advances in Bio and Medical Sciences}}. Springer,
  \bibinfo{pages}{127--141}.
\newblock


\bibitem[\protect\citeauthoryear{Minh et~al\mbox{.}}{Minh
  et~al\mbox{.}}{2020}]%
        {minh_2020_iqtree21}
\bibfield{author}{\bibinfo{person}{Bui~Quang Minh} {et~al\mbox{.}}}
  \bibinfo{year}{2020}\natexlab{}.
\newblock \showarticletitle{IQ-TREE 2: New Models and Efficient Methods for
  Phylogenetic Inference in the Genomic Era}.
\newblock \bibinfo{journal}{\emph{Molecular Biology and Evolution}}
  \bibinfo{volume}{37}, \bibinfo{number}{5} (\bibinfo{year}{2020}),
  \bibinfo{pages}{1530--1534}.
\newblock
\urldef\tempurl%
\url{https://doi.org/10.1093/molbev/msaa015}
\showDOI{\tempurl}


\bibitem[\protect\citeauthoryear{Naveca et~al\mbox{.}}{Naveca
  et~al\mbox{.}}{2021}]%
        {naveca2021phylogenetic}
\bibfield{author}{\bibinfo{person}{F. Naveca} {et~al\mbox{.}}}
  \bibinfo{year}{2021}\natexlab{}.
\newblock \showarticletitle{Phylogenetic relationship of SARS-CoV-2 sequences
  from Amazonas with emerging Brazilian variants harboring mutations E484K and
  N501Y in the Spike protein}.
\newblock \bibinfo{journal}{\emph{Virological. org}}  \bibinfo{volume}{1}
  (\bibinfo{year}{2021}).
\newblock


\bibitem[\protect\citeauthoryear{Ranstam and Cook}{Ranstam and Cook}{2018}]%
        {ranstam2018lasso}
\bibfield{author}{\bibinfo{person}{J Ranstam} {and} \bibinfo{person}{JA Cook}.}
  \bibinfo{year}{2018}\natexlab{}.
\newblock \showarticletitle{LASSO regression}.
\newblock \bibinfo{journal}{\emph{Journal of British Surgery}}
  \bibinfo{volume}{105}, \bibinfo{number}{10} (\bibinfo{year}{2018}),
  \bibinfo{pages}{1348--1348}.
\newblock


\bibitem[\protect\citeauthoryear{{SARS-CoV-2 Variant Classifications and
  Definitions}}{{SARS-CoV-2 Variant Classifications and Definitions}}{2021}]%
        {CDS_variantDef}
\bibfield{author}{\bibinfo{person}{{SARS-CoV-2 Variant Classifications and
  Definitions}}.} \bibinfo{year}{2021}\natexlab{}.
\newblock
  \bibinfo{howpublished}{\url{https://www.cdc.gov/coronavirus/2019-ncov/variants/variant-info.html}}.
\newblock
\newblock
\shownote{[Online; accessed 1-September-2021].}


\bibitem[\protect\citeauthoryear{Satopaa, Albrecht, Irwin, and
  Raghavan}{Satopaa et~al\mbox{.}}{2011}]%
        {satopaa2011finding}
\bibfield{author}{\bibinfo{person}{Ville Satopaa}, \bibinfo{person}{Jeannie
  Albrecht}, \bibinfo{person}{David Irwin}, {and} \bibinfo{person}{Barath
  Raghavan}.} \bibinfo{year}{2011}\natexlab{}.
\newblock \showarticletitle{Finding a" kneedle" in a haystack: Detecting knee
  points in system behavior}. In \bibinfo{booktitle}{\emph{International
  conference on distributed computing systems workshops}}. IEEE,
  \bibinfo{pages}{166--171}.
\newblock


\bibitem[\protect\citeauthoryear{Shakeel., Karim, and Khan}{Shakeel.
  et~al\mbox{.}}{2019}]%
        {Shakeel2019MultiBilingual}
\bibfield{author}{\bibinfo{person}{M. Shakeel.}, \bibinfo{person}{A. Karim},
  {and} \bibinfo{person}{I. Khan}.} \bibinfo{year}{2019}\natexlab{}.
\newblock \showarticletitle{A Multi-cascaded Deep Model for Bilingual SMS
  Classification}. In \bibinfo{booktitle}{\emph{International Conference on
  Neural Information Processing (ICONIP)}}. \bibinfo{pages}{287--298}.
\newblock


\bibitem[\protect\citeauthoryear{Shakeel, Karim, and Khan}{Shakeel
  et~al\mbox{.}}{2020b}]%
        {Shakeel2020Multi}
\bibfield{author}{\bibinfo{person}{M. Shakeel}, \bibinfo{person}{A. Karim},
  {and} \bibinfo{person}{I. Khan}.} \bibinfo{year}{2020}\natexlab{b}.
\newblock \showarticletitle{A Multi-Cascaded Model with Data Augmentation for
  Enhanced Paraphrase Detection in Short Texts}.
\newblock \bibinfo{journal}{\emph{Information Processing \& Management}}
  \bibinfo{volume}{57} (\bibinfo{year}{2020}), \bibinfo{pages}{1--19}.
\newblock


\bibitem[\protect\citeauthoryear{Shakeel, Faizullah, Alghamidi, and
  Khan}{Shakeel et~al\mbox{.}}{2020a}]%
        {Shakeel2020LanguageIndependent}
\bibfield{author}{\bibinfo{person}{M.~H. Shakeel}, \bibinfo{person}{S.
  Faizullah}, \bibinfo{person}{T. Alghamidi}, {and} \bibinfo{person}{I. Khan}.}
  \bibinfo{year}{2020}\natexlab{a}.
\newblock \showarticletitle{Language independent sentiment analysis}. In
  \bibinfo{booktitle}{\emph{International Conference on Advances in the
  Emerging Computing Technologies (AECT)}}. \bibinfo{pages}{1--5}.
\newblock


\bibitem[\protect\citeauthoryear{Stephens et~al\mbox{.}}{Stephens
  et~al\mbox{.}}{2015}]%
        {stephens-2015-genomical}
\bibfield{author}{\bibinfo{person}{Z.~D. Stephens} {et~al\mbox{.}}}
  \bibinfo{year}{2015}\natexlab{}.
\newblock \showarticletitle{Big Data: Astronomical or Genomical?}
\newblock \bibinfo{journal}{\emph{PLoS Biology}} (\bibinfo{year}{2015}).
\newblock
\urldef\tempurl%
\url{https://doi.org/10.1371/journal.pbio.1002195}
\showDOI{\tempurl}


\bibitem[\protect\citeauthoryear{Tariq, Ahmad, Khan, and Shabbir}{Tariq
  et~al\mbox{.}}{2017}]%
        {Tariq2017Scalable}
\bibfield{author}{\bibinfo{person}{J. Tariq}, \bibinfo{person}{M. Ahmad},
  \bibinfo{person}{I. Khan}, {and} \bibinfo{person}{M. Shabbir}.}
  \bibinfo{year}{2017}\natexlab{}.
\newblock \showarticletitle{Scalable Approximation Algorithm for Network
  Immunization}. In \bibinfo{booktitle}{\emph{Pacific Asia Conference on
  Information Systems (PACIS)}}. \bibinfo{pages}{200}.
\newblock


\bibitem[\protect\citeauthoryear{Tibshirani}{Tibshirani}{1996}]%
        {tibshirani1996regression}
\bibfield{author}{\bibinfo{person}{Robert Tibshirani}.}
  \bibinfo{year}{1996}\natexlab{}.
\newblock \showarticletitle{Regression shrinkage and selection via the lasso}.
\newblock \bibinfo{journal}{\emph{Journal of the Royal Statistical Society:
  Series B (Methodological)}} \bibinfo{volume}{58}, \bibinfo{number}{1}
  (\bibinfo{year}{1996}), \bibinfo{pages}{267--288}.
\newblock


\bibitem[\protect\citeauthoryear{Ullah, Ali, Khan, Khan, and Faizullah}{Ullah
  et~al\mbox{.}}{2020}]%
        {ullah2020effect}
\bibfield{author}{\bibinfo{person}{A. Ullah}, \bibinfo{person}{S. Ali},
  \bibinfo{person}{I. Khan}, \bibinfo{person}{M.A. Khan}, {and}
  \bibinfo{person}{S. Faizullah}.} \bibinfo{year}{2020}\natexlab{}.
\newblock \showarticletitle{Effect of Analysis Window and Feature Selection on
  Classification of Hand Movements Using EMG Signal}. In
  \bibinfo{booktitle}{\emph{SAI Intelligent Systems Conference (IntelliSys)}}.
  \bibinfo{pages}{400--415}.
\newblock


\bibitem[\protect\citeauthoryear{Van~der M. and Hinton}{Van~der M. and
  Hinton}{2008}]%
        {van2008visualizing}
\bibfield{author}{\bibinfo{person}{L. Van~der M.} {and} \bibinfo{person}{G.
  Hinton}.} \bibinfo{year}{2008}\natexlab{}.
\newblock \showarticletitle{Visualizing data using t-SNE.}
\newblock \bibinfo{journal}{\emph{Journal of Machine Learning Research (JMLR)}}
  \bibinfo{volume}{9}, \bibinfo{number}{11} (\bibinfo{year}{2008}).
\newblock


\bibitem[\protect\citeauthoryear{Wu, Zhao, Yu, Chen, Wang, Song, Hu, Tao, Tian,
  Pei, et~al\mbox{.}}{Wu et~al\mbox{.}}{2020}]%
        {wu2020new}
\bibfield{author}{\bibinfo{person}{Fan Wu}, \bibinfo{person}{Su Zhao},
  \bibinfo{person}{Bin Yu}, \bibinfo{person}{Yan-Mei Chen},
  \bibinfo{person}{Wen Wang}, \bibinfo{person}{Zhi-Gang Song},
  \bibinfo{person}{Yi Hu}, \bibinfo{person}{Zhao-Wu Tao},
  \bibinfo{person}{Jun-Hua Tian}, \bibinfo{person}{Yuan-Yuan Pei},
  {et~al\mbox{.}}} \bibinfo{year}{2020}\natexlab{}.
\newblock \showarticletitle{A new coronavirus associated with human respiratory
  disease in China}.
\newblock \bibinfo{journal}{\emph{Nature}} \bibinfo{volume}{579},
  \bibinfo{number}{7798} (\bibinfo{year}{2020}), \bibinfo{pages}{265--269}.
\newblock


\bibitem[\protect\citeauthoryear{Yadav et~al\mbox{.}}{Yadav
  et~al\mbox{.}}{2021}]%
        {yadav2021neutralization}
\bibfield{author}{\bibinfo{person}{P. Yadav} {et~al\mbox{.}}}
  \bibinfo{year}{2021}\natexlab{}.
\newblock \showarticletitle{Neutralization potential of Covishield vaccinated
  individuals sera against B. 1.617. 1}.
\newblock \bibinfo{journal}{\emph{bioRxiv}}  \bibinfo{volume}{1}
  (\bibinfo{year}{2021}).
\newblock


\bibitem[\protect\citeauthoryear{Zhang et~al\mbox{.}}{Zhang
  et~al\mbox{.}}{2021}]%
        {zhang2021emergence}
\bibfield{author}{\bibinfo{person}{W. Zhang} {et~al\mbox{.}}}
  \bibinfo{year}{2021}\natexlab{}.
\newblock \showarticletitle{Emergence of a novel SARS-CoV-2 variant in Southern
  California}.
\newblock \bibinfo{journal}{\emph{{Jama}}} \bibinfo{volume}{325},
  \bibinfo{number}{13} (\bibinfo{year}{2021}), \bibinfo{pages}{1324--1326}.
\newblock


\bibitem[\protect\citeauthoryear{Zielezinski, Vinga, Almeida, and
  Karlowski}{Zielezinski et~al\mbox{.}}{2017}]%
        {Zielezinski2017AlignmentfreeSC}
\bibfield{author}{\bibinfo{person}{A. Zielezinski}, \bibinfo{person}{S. Vinga},
  \bibinfo{person}{J. Almeida}, {and} \bibinfo{person}{W. Karlowski}.}
  \bibinfo{year}{2017}\natexlab{}.
\newblock \showarticletitle{Alignment-free sequence comparison}.
\newblock \bibinfo{journal}{\emph{Genome Biology}}  \bibinfo{volume}{18}
  (\bibinfo{year}{2017}).
\newblock


\end{thebibliography}

\end{document}